# Efficient Inference in Markov Control Problems

**Thomas Furmston**
Computer Science Department
University College London
London, WC1E 6BT

**David Barber**
Computer Science Department
University College London
London, WC1E 6BT

## Abstract

Markov control algorithms that perform smooth, non-greedy updates of the policy have been shown to be very general and versatile, with policy gradient and Expectation Maximisation algorithms being particularly popular. For these algorithms, marginal inference of the reward weighted trajectory distribution is required to perform policy updates. We discuss a new exact inference algorithm for these marginals in the finite horizon case that is more efficient than the standard approach based on classical forward-backward recursions. We also provide a principled extension to infinite horizon Markov Decision Problems that explicitly accounts for an infinite horizon. This extension provides a novel algorithm for both policy gradients and Expectation Maximisation in infinite horizon problems.

## 1 MARKOV DECISION PROBLEMS

A Markov Decision Problem (MDP) is described by an initial state distribution $p_1(s_1)$, transition distributions $p(s_{t+1}|s_t, a_t)$ and reward function $R_t(s_t, a_t)$, where the state and action at time $t$ are denoted by $s_t$ and $a_t$ respectively[1] (Sutton and Barto, 1998). The state and action spaces can be either discrete or continuous. For a discount factor $\gamma$ the reward is defined as $R_t(s_t, a_t) = \gamma^{t-1} R(s_t, a_t)$ for a stationary reward $R(s_t, a_t)$, where $\gamma \in [0, 1)$. We assume a stationary policy, $\pi$, defined as a set of conditional distributions over the action space, $\pi_{a,s} = p(a_t = a|s_t = s, \pi)$. The total expected reward of the MDP (the policy utility) is given by

$$U(\pi) = \sum_{t=1}^{H} \sum_{s_t, a_t} R_t(s_t, a_t) p(s_t, a_t|\pi) \quad (1)$$

where $H$ is the horizon, which can be either finite or infinite, and $p(s_t, a_t|\pi)$ is the marginal of the joint state-action trajectory distribution

$$p(s_{1:H}, a_{1:H}|\pi) = p(a_H|s_H, \pi) p_1(s_1) \\ \times \prod_{t=1}^{H-1} p(s_{t+1}|s_t, a_t) p(a_t|s_t, \pi). \quad (2)$$

Given a transition model $p(s_{t+1}|s_t, a_t)$, the MDP learning problem is to find a policy $\pi$ that maximises (1). For all but the most select cases, such as small discrete environments or linear-quadratic control, this is a notoriously difficult optimisation problem which has given rise to a multitude of competing approaches.

Classical planning algorithms, such as Policy Iteration or Value Iteration (Sutton and Barto, 1998), generally focus on greedy updates of the policy. While these algorithms work well in small discrete environments it has been difficult to extend them to more complex problems, such as structured or continuous domains. Additionally, these greedy updates of the policy become increasingly unstable as the problem domain becomes more complex. As a consequence a large amount of research has been done in designing algorithms which perform smooth policy updates. Gradient ascent algorithms, *e.g.* (Sutton et al., 2000), and the Expectation Maximisation algorithm, *e.g.* (Toussaint et al., 2006), have been particularly popular and have been successfully applied to a large range of complex domains including optimal control (Toussaint et al., 2006), robotics (Kober and Peters, 2009; Peters and Schaal, 2006) and Bayesian reinforcement learning (Furmston and Barber, 2010).

---

[1] To avoid cumbersome notation we also use the notation $z_t = \{s_t, a_t\}$ to denote a state-action pair. We use the bold typeface, $\mathbf{z}_t$, to denote a vector.

## 1.1 EXPECTATION MAXIMISATION

By expressing the objective function (1) as the likelihood function of an appropriately constructed mixture model the MDP can be solved using techniques from probabilistic inference, such as EM (Dayan and Hinton, 1997; Toussaint et al., 2006; Kober and Peters, 2009; Toussaint et al., 2011), EP (Toussaint, 2009; Furmston and Barber, 2010) or MCMC (Hoffman et al., 2008). Without loss of generality, we assume the reward is non-negative and define the reward weighted path distribution

$$\hat{p}(s_{1:t}, a_{1:t}, t|\pi) = \frac{R_t(s_t, a_t)p(s_{1:t}, a_{1:t}|\pi)}{U(\pi)}. \quad (3)$$

This distribution is properly normalised, as can be seen from (1) and (2). The graphical structure of this distribution is given by a set of chains, each corresponding to a different time-point at which a reward is received, see figure 1 for an example.

We now define a variational distribution $q(s_{1:t}, a_{1:t}, t)$, and take the Kullback-Leibler divergence between the $q$-distribution and (3). Since

$$\mathrm{KL}(q(s_{1:t}, a_{1:t}, t)||\hat{p}(s_{1:t}, a_{1:t}, t|\pi)) \geq 0 \quad (4)$$

we obtain a lower bound on the log utility

$$\log U(\pi) \geq \tilde{H}(q(s_{1:t}, a_{1:t}, t)) + \langle \log \tilde{p}(s_{1:t}, a_{1:t}, t|\pi) \rangle_q \quad (5)$$

where $\langle \cdot \rangle_q$ denotes the average w.r.t. $q(s_{1:t}, a_{1:t}, t)$, $\tilde{H}(\cdot)$ is the entropy function and $\tilde{p}$ is the unnormalised reward weighted trajectory distribution. An EM algorithm can be obtained from the bound in (5) by iterative coordinate-wise maximisation:

**E-step** For fixed $\pi^{old}$ find the best $q$ that maximises the r.h.s. of (5). For no constraint on $q$, this gives $q = \hat{p}(s_{1:t}, a_{1:t}, t|\pi^{\mathrm{old}})$. Then compute $\hat{p}(s_\tau = s, a_\tau = a, t|\pi^{\mathrm{old}})$, the marginals of the reward weighted path distribution equation (3) evaluated at the previous policy.

**M-step** For fixed $q$ find the best $\pi$ that maximises the r.h.s. of (5). This is equivalent to maximising w.r.t. $\pi$ the 'energy' contribution

$$\sum_{t=1}^{H}\sum_{\tau=1}^{t} \langle \log \pi(a_\tau|s_\tau) \rangle_{\hat{p}(s_\tau, a_\tau, t|\pi^{\mathrm{old}})} \quad (6)$$

Note that for an unconstrained tabular policy, the M-step gives the policy update

$$\pi(s, a) \propto \sum_{t=1}^{H}\sum_{\tau=1}^{t} \hat{p}(s_\tau = s, a_\tau = a, t|\pi^{\mathrm{old}}). \quad (7)$$

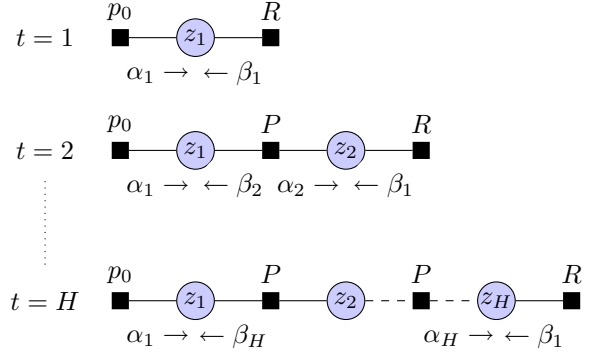

Figure 1: An example of the forward-backward recursions that are performed in finite horizon EM algorithm on the corresponding factor graphs.

## 1.2 POLICY GRADIENTS

The policy gradients algorithm iteratively updates the policy parameters in the direction of the gradient of $\nabla_\pi U(\pi)$, in order to increase $U(\pi)$ and thereby improve the policy. These gradients can be computed using the identity, see e.g. (Salakhutdinov et al., 2003),

$$\partial_\pi \log U(\pi) = \sum_{t=1}^{H}\sum_{\tau=1}^{t} \langle \partial_\pi \log \pi(a_\tau|s_\tau) \rangle_{\hat{p}(s_\tau, a_\tau, t|\pi)}. \quad (8)$$

To perform a policy update both the EM and policy gradient algorithms therefore require sufficient statistics of the reward weighted path distribution in equation (3); either the state-action marginals in discrete problems or the moments in exponential family continuous problems.

We note that in terms of inference the only difference between these two algorithms is that in the policy gradients algorithm the reward weighted distribution depends on the current policy, while in the EM algorithm it is dependent on the policy of the previous EM step. To ease notation we denote the distribution (3) of both algorithms by $q(z_{1:t}, t)$, see footnote 1.

## 1.3 FORWARD-BACKWARD INFERENCE

To perform a policy update for finite horizon problems we need to calculate the marginals $q(z_\tau, t)$, $\forall t \in \{1, \ldots, H\}$ and $\tau \in \{1, \ldots, t\}$. For each component, $t$, the distribution $q(z_{1:t}|t)$ is chain structured and so all the marginals $q(z_\tau|t)$ can be calculated in linear time using message-passing, see e.g. (Wainwright and Jordan, 2008). In particular these marginals can be calculated using forward-backward recursions, otherwise known as $\alpha$-$\beta$ recursions. The initial messages,

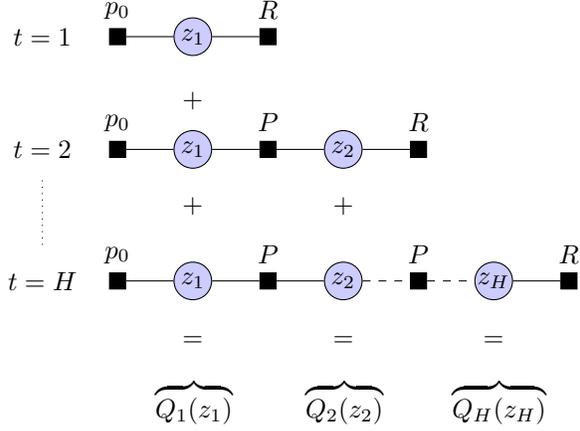

Figure 2: An example of how the finite horizon reward weighted trajectory distribution (3) splits into $Q$-functions.

$\alpha_1$, $\beta_1$, are given by

$$\alpha_1(z) = p_0(s)p(a|s;\pi), \qquad \beta_1(z) = R(s,a),$$

and the forward-backward recursions are given by

$$\alpha_{\tau+1}(z') = P(z'|z)\alpha_\tau(z), \quad \beta_{\tau+1}(z) = P(z'|z)^T \beta_\tau(z')$$

where $P(z'|z)$ is the state-action transition distribution, which is given by $P(z'|z) = p(a'|s';\pi)p(s'|s,a)$. Given the sets of forward-backward messages each individual state-action marginal is given

$$q(z_\tau = z|t) \propto \alpha_\tau(z)\beta_{t+1-\tau}(z).$$

The marginals can be calculated concurrently through reuse of the forward-backward messages; see figure 1 for an illustration of the reuse of the forward-backward recursions. As there are $H$ separate time components, where each component, $t$, has $t$ state-action marginals the total computational complexity of calculating the policy update equation is $\mathcal{O}(H^2)$.

## 2 Q-INFERENCE

We proceed to show that the previously described $\mathcal{O}(H^2)$ forward-backward algorithm doesn't fully exploit the conditional independence structure of the distribution (3) and that a more efficient $\mathcal{O}(H)$ procedure exists. We focus first on the finite horizon case, for which an exact algorithm exists, before extending the approach to the infinite horizon in §2.1.

We first prove the following lemma which shows that, conditioned on the succeeding state-action pair $z_{\tau+1}$ and the time-component $t$, the reward weighted trajectory distribution over $z_\tau$ is independent of $t$ and is equal to the system reversal dynamics.

**Lemma 1.** *Given $t, \tau \in \{1, \ldots, H-1\}$ s.t. $\tau < t$, then the variational distribution $q(z_\tau|z_{\tau+1}, t)$ is independent of $t$ and takes the form*

$$q(z_\tau|z_{\tau+1}, t) = p(z_\tau|z_{\tau+1}) \tag{9}$$

*where $p(z_\tau|z_{\tau+1})$ is the marginal of the trajectory distribution (2).*

*Proof.* For any given $\tau \in \{1, \ldots, t\}$ the marginal of the variational distribution $q(z_{\tau:t}|t)$ takes the form

$$q(z_{\tau:t}|t) = \frac{1}{V_t} p(z_\tau) \bigg\{ \prod_{\tau'=\tau}^{t-1} p(z_{\tau'+1}|z_{\tau'}) \bigg\} R(z_t),$$

where $V_t = \mathbb{E}[R(z_t)]$ is the normalisation constant. As $\tau < t$ we have a similar expression for the marginal $q(z_{\tau+1:t}|t)$. Using the Markovian structure of the variational distribution means that the conditional distribution takes the form

$$q(z_\tau|z_{\tau+1}, t) = \frac{p(z_\tau)p(z_{\tau+1}|z_\tau)}{p(z_{\tau+1})} = p(z_\tau|z_{\tau+1}).$$

$\square$

We now introduce a new set of '$Q$-functions' that will play a prominent role in the rest of the paper. For each $\tau \in \{1, \ldots, H\}$ we define the function

$$Q_\tau(z) = \sum_{t=\tau}^{H} q(z_\tau = z, t). \tag{10}$$

Note that the sum of the state-action marginals in both the policy gradient equation and the EM policy update function can be written in terms of the $Q$-functions as follows

$$\sum_{t=1}^{H}\sum_{\tau=1}^{t} q(z_\tau, t) = \sum_{\tau=1}^{H}\sum_{t=\tau}^{H} q(z_\tau, t) = \sum_{\tau=1}^{H} Q_\tau(z_\tau). \tag{11}$$

An illustration of how the marginals of the reward weighted trajectory distribution can be written in terms of the $Q$-functions is given in figure 2.

These $Q$-functions have the intuitive interpretation of being proportional to the probability of reaching state-action pair $z$ at time $\tau$ times the total expected future reward. Therefore the part of the trajectory before time-point $\tau$ plays a prominent role in these functions, which isn't the case in classical planning algorithms.

We now use lemma 1 to obtain a recursive relationship over the $Q$-functions.

**Lemma 2.** *Given $\tau \in \{1, \ldots, H-1\}$, the function $Q_\tau(z)$ satisfies*

$$Q_\tau(z_\tau) = q(z_\tau, \tau) + \sum_{z_{\tau+1}} p(z_\tau|z_{\tau+1}) Q_{\tau+1}(z_{\tau+1}). \tag{12}$$

*Proof.* We start by rewriting the function $Q_\tau(z)$ as

$$Q_\tau(z_\tau) = q(z_\tau, \tau) + \sum_{t=\tau+1}^{H} \sum_{z_{\tau+1}} q(z_\tau, z_{\tau+1}|t)q(t), \quad (13)$$

where we have exploited that fact that $q(\cdot|t)$ is a distribution and introduced the state-action variable of the next time-step, $z_{\tau+1}$. Now, by lemma 1, we have that for each $t \in \{\tau+1, \ldots, H\}$

$$\begin{aligned} q(z_\tau, z_{\tau+1}|t) &= q(z_\tau|z_{\tau+1}, t)q(z_{\tau+1}|t) \\ &= p(z_\tau|z_{\tau+1})q(z_{\tau+1}|t). \end{aligned} \quad (14)$$

Substituting this into (13) we obtain

$$Q_\tau(z_\tau) = q(z_\tau, \tau) + \sum_{z_{\tau+1}} p(z_\tau|z_{\tau+1}) \sum_{t=\tau+1}^{H} q(z_{\tau+1}|t)q(t),$$

where we have used the fact that $p(\boldsymbol{z}_\tau|\boldsymbol{z}_{\tau+1})$ depends only upon $\tau$ and not upon $t$. The result now follows from the definition of $Q_{\tau+1}(z)$. $\square$

We now briefly describe how the recursive relation in lemma 2 can be used to calculate the $Q$-functions in linear time for finite planning horizons. Firstly, the trajectory distribution (2) is chain structured so all the state-action marginals, $\{p(z_\tau)\}_\tau$, can be calculated in linear time. Additionally, the normalisation constant of the distribution (3) is by definition $U(\pi)$ which may also be calculated in linear time by a standard forward recursion. For each $\tau \in \{1, \ldots, H\}$ the term $q(z_\tau = z, \tau)$ that occurs in the function $Q_\tau(z)$ takes the form

$$q(z_\tau, \tau) = \frac{1}{U(\pi)} p(z_\tau) R(z_\tau). \quad (15)$$

Using the previous remarks it is clear that each of these terms can be calculated in linear time (Wainwright and Jordan, 2008). We now make the observation that $Q_H(z_H) = q(z_H, H)$, so that this function can be calculated in linear time. Once the function $Q_H(z_H)$ has been calculated all of the remaining functions $\{Q_\tau(z_\tau)\}_{\tau=1}^{H-1}$ can be computed by repeated use of the recursion (12). There are $H-1$ applications of the recursion, each of which takes a constant amount of time to compute. Once all the $Q$-functions have been calculated a policy update can now be performed through (11).

Before extending our inference algorithm to infinite horizons it is of interest to note the difference between our inference algorithm and the standard forward-backward algorithm. The standard forward-backward algorithm focuses on performing the forward and backward messages concurrently, exploiting the similarity in the components of the reward weighted trajectory distribution to reuse the messages. On the other hand the $Q$-inference algorithm first computes the forward messages and then, using them, performs the backward iterations of the $Q$-functions (12). This enables $Q$-recursions to use the time-independent Markovian structure of the system reversal dynamics in the reward weighted trajectory distribution. This property enables the $Q$-recursions to propagate the effect the current state-action pair has on all future rewards in a single calculation. This is equivalent to performing inference over all future time components of the reward weighted trajectory distribution concurrently.

**Algorithm 1** Finite Horizon Inference of Reward Trajectory Distribution

**Calculate Forward Messages:** Iterate the forward message recursion until the final forward message, $\alpha_H(z)$, has been calculated.
**Calculate the Final $Q$-function:** Use the forward message, $\alpha_H(z)$, and the total expected reward, $U(\pi)$, to calculate the $Q$-function for the final time-point, $Q_H(z)$, using equation (15).
**Calculate Backward Messages:** Use the recursive equation (12) to propagate the $Q$-functions backwards in time $Q_t(z)$, for $t = H-1, \ldots, 1$.

## 2.1 INFINITE PLANNING HORIZON

While we have only considered finite horizon problems the recursive relation of the $Q$-functions can also be used in infinite horizon problems. To do so we rely on the fact that, given the system is controllable, it will reach its stationary state-action distribution in a finite amount of time. Given that the system reaches its stationary state-action distribution by the time-point $\hat{\tau}$, then it is straightforward to show that for any $\tau \geq \hat{\tau}$ we have the relation

$$Q_{\tau+1}(z) = \gamma Q_\tau(z). \quad (16)$$

This relation can now be used to obtain a formulation for calculating the infinite number of state-action marginals of the reward weighted trajectory distribution. Firstly we split the infinite summation in the policy update function into the terms before and after the stationary state-action distribution has been reached,

$$\sum_{t=1}^{\infty} Q_t(z) = \sum_{t=1}^{\hat{\tau}-1} Q_t(z) + \sum_{t=\hat{\tau}}^{\infty} Q_t(z). \quad (17)$$

We now introduce the function $Q(z)$, which is defined by $Q(z) = \gamma^{1-\hat{\tau}} Q_{\hat{\tau}}(z)$. Note that by (16) we have $Q(z) = \gamma^{1-t} Q_t(z)$ for all $t \geq \hat{\tau}$. The infinite summa-

> **Algorithm 2** Infinite Horizon Inference of Reward Trajectory Distribution
>
> **Calculate Forward Messages:** Iterate the forward message recursion until the forward messages converge to the stationary distribution.
> **Calculate Stationary $Q$-function:** Use the stationary state-action distribution and the stationary system reversal dynamics to calculate the stationary $Q$ function, $Q(z)$, using either (20) or the fixed-point equation (19).
> **Calculate Backward Messages:** Use the recursive equation (12) to propagate the $Q$-functions backwards in time $Q_t(z)$, for $t = \hat{\tau} - 1, \ldots, 1$.

tion that occurs (17) can now be performed analytically as follows

$$\sum_{t=\hat{\tau}}^{\infty} Q_t(z) = Q(z) \sum_{t=\hat{\tau}}^{\infty} \gamma^{t-1} = \frac{\gamma^{\hat{\tau}-1}}{1-\gamma} Q(z). \qquad (18)$$

To perform the summation in (17) it now remains to obtain an analytic solution to $Q(z)$. This is obtained from the following recursion, which is easy to prove using the relations (12) and (16) and the definition of $Q(z)$,

$$Q(z) = \frac{\alpha(z)R(z)}{U(\pi)} + \gamma \sum_{z'} \overleftarrow{p}(z|z') Q(z'), \qquad (19)$$

where $\alpha(z)$ is the stationary state-action distribution and $\overleftarrow{p}(z|z')$ is the stationary system reversal dynamics. An algebraic solution for $Q(z)$ is obtained from (19) by observing that

$$Q = (I - \gamma \overleftarrow{P})^{-1} \boldsymbol{\mu}, \qquad (20)$$

where $\boldsymbol{\mu}$ is the point-wise product of the stationary state-action distribution with the reward function scaled by the inverse of the total expected reward. An alternative solution to $Q(z)$ can be obtained by iterating the fixed point equation (19) until convergence, which may be preferable in systems where the matrix inversion is expensive. The complete algorithm for calculating the infinite number of state-action marginals of the reward weighted trajectory distribution is summarized in algorithm (2).

We also note that this gives a new formulation of the infinite horizon EM algorithm. While the current infinite horizon EM algorithm (Toussaint et al., 2006, 2011) uses the idea of the 'time-marginal' to select a finite horizon to approximate the infinite horizon problem, our formulation relies on the convergence of the state-action distribution $p(z)$ to its stationary distribution. Once converged we can use (20) or (19) to calculate the infinite number of marginals needed to perform a policy update.

Before proceeding we make a brief note about the 'time-marginal' criterion used in (Toussaint et al., 2006). It is not possible to implement the infinite horizon policy update function of (Toussaint et al., 2006) exactly and a finite horizon is therefore selected which will give a good approximation[2]. To select an appropriate finite horizon (Toussaint et al., 2006) propose to use the 'time-marginal' $q(t)$, which can be calculated up to proportionality using the equation

$$q(t) \propto \sum_z \alpha_\tau(z) \beta_{t-\tau}(z), \text{ for some } \tau \in \{1, \ldots, t\}.$$

It is proposed that by concurrently iterating forward and backward messages it is a reasonable heuristic to cut-off calculations when $q(t+1) \ll \sum_{\tau=1}^{t} q(\tau)$. We note that this 'time-marginal' is actually the proportion of the objective function that is obtained at the $t^{th}$ time-step, i.e.

$$q(t) = \frac{\mathbb{E}[R(z_t)]}{U(\pi)}.$$

Taking this into account one can expect the 'time-marginal' criterion to perform poorly in situations where the reward function has a sparse multi-modal structure.

## 3 DYNAMIC PROGRAMMING

Equations (12) and (19) bear a strong resemblance to *policy evaluation* from classical infinite horizon planning algorithms, see *e.g.* (Sutton and Barto, 1998),

$$Q^\pi(s,a) = R(s,a) + \gamma \sum_{s',a'} \pi(a'|s') p(s'|s,a) Q^\pi(s',a').$$

However, while there is a strong resemblance there are also some significant differences. Firstly note that the $Q$-functions (12) and (19) are weighted inversely by the total expected reward. This occurs because EM and policy gradients work in probability space. Additionally, the standard $Q$-functions of policy evaluation represent the total expected future reward given the *current* state-action pair, and so do not depend on any previous time-steps. This is in contrast to the $Q$-functions of (12) and (19) which explicitly depend on previous time-steps through the state-action marginal. The reason for this difference can be explained through the different nature of the policy updates of these algorithms. The policy update equation of *policy improvement* takes the greedy form

$$\pi^{\text{new}}(a|s) = \delta_{a,a^*(s)}, \quad a^*(s) = \operatorname*{argmax}_{a} Q^\pi(s,a).$$

---

[2]We have recently become of aware of a new formulation of the EM algorithm, called *incremental EM* (Toussaint et al., 2011), that also converges in the limit. This incremental EM has a convergence rate that is exponential in the discount factor $\gamma$, while our algorithm has a convergence rate that is exponential in the magnitude of the largest eigenvalue (excluding eigenvalues equal to unity) of the state-action transition matrix. Additionally, our methods generalise to other algorithms that are based on the reward weighted trajectory distribution.

As an MDP is chain structured the maximisation over $a$ given $s$ is independent of any of the previous time-steps, meaning that these $Q$-functions need only depend on future time-points. Meanwhile, the EM and policy gradients algorithms don't condition on the state variable in this way during the policy update and so this splitting of the chain structure doesn't occur.

Finally, in the finite horizon case it can be seen through (11) that there corresponds $H$ different $Q$-functions to the policy update (7) or the policy gradient (8), whereas in dynamic programming there is a one to one correspondence. This is because the algorithms highlighted in §1.1 and §1.2 are for stationary policies. Either of these algorithms can easily be re-derived for non-stationary policies and this will result in a one to one correspondence between the $Q$-functions and the non-stationary policies, *i.e.* $Q_t$ will correspond to $\pi_t$.

## 4 CONTINUOUS MDPs

The proofs of lemma 1 and lemma 2 follow over to the continuous case easily and the continuous version of equation (12) takes the form

$$Q_\tau(\boldsymbol{z}_\tau) = q(\boldsymbol{z}_\tau, \tau) + \int d\boldsymbol{z}_{\tau+1} p(\boldsymbol{z}_\tau|\boldsymbol{z}_{\tau+1}) Q_{\tau+1}(\boldsymbol{z}_{\tau+1}). \tag{21}$$

Due to the summation in (21) we can see that the $Q$-functions take the form of a two component mixture model, with one component corresponding to the immediate reward while the second corresponds to future rewards. Although one could model this mixture explicitly it is generally only necessary to calculate the moments

$$\boldsymbol{\mu}^i = \int \boldsymbol{z}^i Q_\tau(\boldsymbol{z}) d\boldsymbol{z}, \quad i \in \{1, \ldots, N\}, N \in \mathbb{N}$$

to perform a policy update. The recursive equation (21) can then be used to calculate these moments recursively in linear time, as we now demonstrate.

For illustrative purposes we consider the specific example of linear continuous MDPs with arbitrary rewards (Hoffman et al., 2009). In this problem class the initial state distribution, policy and transition distributions are linear, with Gaussian noise, and take the parametric form

$$p(s_1) = \mathcal{N}(s_1|\mu_0, \Sigma_0),$$
$$p(s_{t+1}|s_t, a_t) = \mathcal{N}(s_{t+1}|As_t + Ba_t, \Sigma),$$
$$p(a_t|s_t; K, m, \pi_\sigma) = \mathcal{N}(a_t|Ks_t + m; \pi_\sigma),$$

where $\mathcal{N}(x|\boldsymbol{\mu}, \Sigma)$ is a Gaussian distribution with mean and covariance $\boldsymbol{\mu}$ and $\Sigma$ respectively. Additionally, the reward function is a mixture of unnormalised Gaussian distributions[3]

$$R(z_t) = \sum_{j=1}^{J} w_j \bar{\mathcal{N}}(\boldsymbol{y}_j|M\boldsymbol{z}_t, L_j),$$

where $\bar{\mathcal{N}}$ denotes an unnormalised Gaussian. In our $Q$-inference algorithm the first step is to calculate the forward messages, which is achieved using standard forward message recursions, see *e.g.* (Hoffman et al., 2009),

$$\boldsymbol{\mu}_{t+1} = F\boldsymbol{\mu}_t + \bar{\boldsymbol{m}}, \quad \Sigma_{t+1} = F\Sigma_t F^T + \bar{\Sigma}, \quad \bar{\boldsymbol{m}} = \begin{bmatrix} 0 \\ \boldsymbol{m} \end{bmatrix},$$

where $\boldsymbol{\mu}_t$ and $\Sigma_t$ denote the mean and covariance of $\alpha_t(\boldsymbol{z})$ respectively, and

$$\bar{\Sigma} = \begin{bmatrix} \Sigma & \Sigma K^T \\ K\Sigma & K\Sigma K^T + \pi_\sigma I_{n_u} \end{bmatrix}, \quad F = \begin{bmatrix} A & B \\ KA & KB \end{bmatrix}.$$

As the system is linear the reversal dynamics can be calculated using standard conditional Gaussian formulae, see *e.g.* (Barber, 2011). Given the forward messages the system reversal dynamics are given by

$$p(\boldsymbol{z}_t|\boldsymbol{z}_{t+1}) = \mathcal{N}(\boldsymbol{z}_t|G_t\boldsymbol{z}_{t+1} + \overleftarrow{\boldsymbol{m}}_t, \overleftarrow{\Sigma}_t),$$

where $G_t$, $\overleftarrow{\boldsymbol{m}}_t$ and $\overleftarrow{\Sigma}_t$ are given by

$$G_t = \Sigma_t F^T (F\Sigma_t F^T + \bar{\Sigma})^{-1},$$
$$\overleftarrow{\boldsymbol{m}}_t = \boldsymbol{\mu}_t - G_t(F\boldsymbol{\mu}_t + \bar{\boldsymbol{m}}),$$
$$\overleftarrow{\Sigma}_t = \Sigma_t - \Sigma_t F^T (F\Sigma_t F^T + \bar{\Sigma}) F\Sigma_t.$$

We denote the first two moments of $Q_t(\boldsymbol{z})$ by $\boldsymbol{\mu}_t^Q$ and $\Sigma_t^Q$ respectively. Similarly we denote the first two moments of $q(\boldsymbol{z}_t|t)$ by $\boldsymbol{\mu}_t^R$ and $\Sigma_t^R$ respectively. As $\boldsymbol{z}_t$ depends on $\boldsymbol{z}_{t+1}$ linearly in the reversal dynamics it means that the first moment $\int_{\boldsymbol{z}_t} \boldsymbol{z}_t Q_t(\boldsymbol{z}_t)$ takes the form

$$\int_{\boldsymbol{z}_t} \boldsymbol{z}_t q(\boldsymbol{z}_t, t)$$
$$+ \int_{\boldsymbol{z}_t, \boldsymbol{z}_{t+1}} (\boldsymbol{z}_t + G_t \boldsymbol{z}_{t+1} + \overleftarrow{\boldsymbol{m}}_t) \mathcal{N}(\boldsymbol{z}_t|0, \overleftarrow{\Sigma}_t) Q_{t+1}(\boldsymbol{z}_{t+1}),$$

---
[3]For simplicity of exposition we restrict our attention to the case $J = 1$, as the extension to $J > 1$ is trivial.

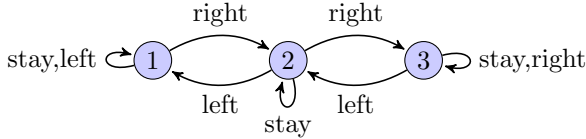

Figure 3: An example of the transition dynamics for the double reward chain of chain problem with 3 states.

with a similar formula for the second moment. Defining, $Z_{t+1} = \sum_{\tau=t+1}^{H} q(\tau)$, and using $\int_{z_{t+1}} Q_{t+1}(z_{t+1}) = Z_{t+1}$, we obtain recursions for the first two moments of the $Q$-functions,

$$\boldsymbol{\mu}_t^Q = q(t)\boldsymbol{\mu}_t^R + Z_{t+1}\overleftarrow{\boldsymbol{m}}_t + G_t\boldsymbol{\mu}_{t+1}^Q \qquad (22)$$

$$\boldsymbol{\Sigma}_t^Q = q(t)\Sigma_t^R + Z_{t+1}(\overleftarrow{\Sigma}_t + \overleftarrow{\boldsymbol{m}}_t\overleftarrow{\boldsymbol{m}}_t^T) \\ + G_t(\Sigma_{t+1}^Q + \boldsymbol{\mu}_{t+1}^Q\overleftarrow{\boldsymbol{m}}_t^T + \overleftarrow{\boldsymbol{m}}_t(\boldsymbol{\mu}_{t+1}^Q)^T)G_t^T. \qquad (23)$$

Given these moments the policy is updated by first solving a set of linear equations in $K$ and $m$, and then solving for $\pi_\sigma$, see (Hoffman et al., 2009).

To summarize, instead of calculating the forward and backward messages concurrently and then calculating the marginals $q(z_\tau, t)$ separately, as in (Hoffman et al., 2009), we have first calculated the forward messages and then used (22) and (23) to calculate the moments of the $Q$-functions recursively. These recursive equations allow the moments necessary for a policy update to be calculated in linear time, which compares favorably with the forward-backward recursions of (Hoffman et al., 2009) that have quadratic runtime.

## 5 EXPERIMENTS

### 5.1 INFINITE HORIZON MDPs

The first experiment we performed was on the *double reward chain problem*, which was designed to highlight the susceptibility of the infinite horizon EM algorithm to get caught in local optima when the 'time-marginal' criterion (Toussaint et al., 2006) is used as a convergence criterion and the 'time-marginal' is multi-modal.

The $N$ length *double reward chain problem* has $N$ states, where we label the states from left to right in the chain, see figure 3. In each state there are three possible actions; to move *left* in the chain, to move *right* in the chain or to *stay* in the current state. When the agent is in the left end-point of the chain and moves *left* it remains in the same state, with a similar situation in the right end-point. All the transition dynamics of the system are deterministic. The agent receives a reward for *staying* in either of the two end-points of the chain. This means there are two optimal types of behaviour: to move towards the left end-point or to move to the right end-point, which we denote by $\pi^{\text{left}}$ and $\pi^{\text{right}}$ respectively. The global/local optimality of these two policies depends on the initial state distribution and the reward function. We defined the initial state as the state adjacent to the left end-point of the chain, while the reward function was defined by

$$R(s = 1, a = stay) = \gamma^{-1},$$
$$R(s = N, a = stay) = 20\gamma^{2-N}.$$

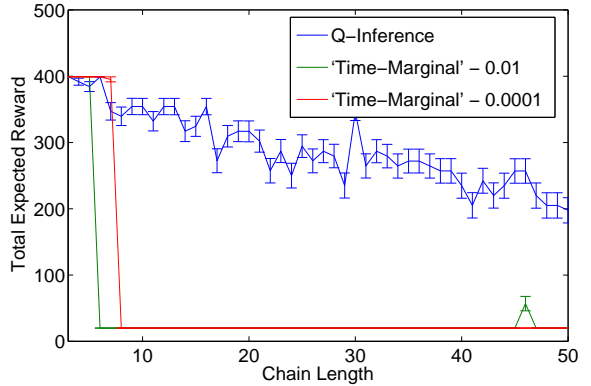

Figure 4: Average total expected reward against the length of the chain for the infinite horizon algorithm §2.1 (blue) and the 'time-marginal' infinite horizon algorithm with $\eta = 0.01$ (green), $\eta = 0.0001$ (red).

The reward function was designed so that the total expected reward of the two optimal policies remains the same regardless of the length of the chain, *i.e.*

$$U(\pi^{left}) = 20, \qquad U(\pi^{right}) = 400.$$

It is therefore always optimal to move towards the right end-point of the chain.

While $\pi^{right}$ is always optimal the initial state of the agent is adjacent to the left end-point. This means that for sufficiently large $N$ the 'time marginal' is likely to have a sparse multi-modal structure, with one mode around low values of $t$ (corresponding to the reward from the left end-point) and the second mode around larger values of $t$ (corresponding to the reward from the right end-point). We ran the experiment for increasing values of $N$, with $N = 3, \ldots, 50$, and repeated the experiment 100 times. In our infinite horizon EM algorithm we propagated forward messages until the change in magnitude of the forward messages didn't exceed 0.01. For the 'time-marginal' stopping criterion we stopped message-passing when $q(t) \leq \eta \sum_{\tau=1}^{t-1} q(\tau)$, where we set $\eta = 0.01$ and $\eta = 0.0001$. The results of the experiment are shown in figure 4, with our infinite horizon algorithm §2.1 (blue) and the 'time-marginal' infinite horizon algorithm with $\eta = 0.01$ (green) and with $\eta = 0.0001$ (red). As can be seen from figure 4 even for large $N$ our infinite horizon EM algorithm is still able to pick up the global optimum at the right end-point of the chain. This compares favorably with the 'time-marginal' criterion, which for $N \geq 10$ is generally unable to pick up the global optimum and instead only finds the local optimum.

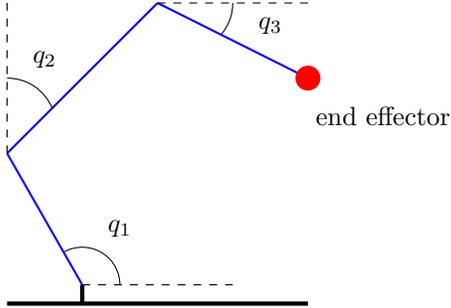

Figure 5: A graphical depiction of a 3-link robot manipulator arm, where the angles of the joints are given by $q_1$, $q_2$ and $q_3$ respectively.

It can also be seen that as $N$ increases the performance of $Q$-inference infinite horizon EM algorithm decreases. This is because the performance of the EM algorithm depends on the initialization of the policy, and as $N$ increases the chance of a favourable initial policy decreases. Parameter initialization is a general problem of the EM algorithm and the standard solution is to make multiple runs with different initializations, selecting the best final result.

## 5.2 ROBOT JOINT MANIPULATOR

The $N$-link rigid robot arm manipulator is a standard continuous model, consisting of an end effector connected to $N$ linked rigid bodies. A graphical depiction of a 3-link rigid manipulator is given in figure 5. A typical continuous control problem for such systems is to apply appropriate torque forces to the joints of the manipulator so as to move the end effector into a desired position. The state of the system is given by $\boldsymbol{q}$, $\dot{\boldsymbol{q}}$, $\ddot{\boldsymbol{q}} \in \mathbb{R}^N$, where $\boldsymbol{q}$, $\dot{\boldsymbol{q}}$ and $\ddot{\boldsymbol{q}}$ denote the angles, velocities and accelerations of the joints respectively, while the control variables are the torques applied to the joints $\boldsymbol{\tau} \in \mathbb{R}^N$. The nonlinear state equations of the system are given by, see *e.g.* (Spong et al., 2005),

$$M(\boldsymbol{q})\ddot{\boldsymbol{q}} + C(\dot{\boldsymbol{q}}, \boldsymbol{q})\dot{\boldsymbol{q}} + g(\boldsymbol{q}) = \boldsymbol{\tau} \tag{24}$$

where $M(\boldsymbol{q})$ is the inertia matrix, $C(\dot{\boldsymbol{q}}, \boldsymbol{q})$ denotes the Coriolis and centripetal forces and $g(\boldsymbol{q})$ is the gravitational force.

While this system is highly nonlinear it is possible to define an appropriate control function $\hat{\boldsymbol{\tau}}(\boldsymbol{q}, \dot{\boldsymbol{q}})$ that results in linear dynamics in a different state-action space. This process is called *feedback linearisation*, see *e.g.* (Khalil, 2001), and in the case of an $N$-link rigid manipulator recasts the torque action space into the acceleration action space. This means that the state of the system is now given by $\boldsymbol{q}$ and $\dot{\boldsymbol{q}}$, while the control is $\boldsymbol{u} = \ddot{\boldsymbol{q}}$.

Ordinarily in such problems the reward would be a function of the generalised co-ordinates of the end effector, which results in a non-trivial reward function in terms of $\boldsymbol{q}$, $\dot{\boldsymbol{q}}$ and $\ddot{\boldsymbol{q}}$. While this reward function can be modelled as a mixture of Gaussians, see (Hoffman et al., 2009), for simplicity we consider the simpler problem where the reward is a function of $\boldsymbol{q}$, $\dot{\boldsymbol{q}}$ and $\ddot{\boldsymbol{q}}$ directly.

In the experiments we considered a 2-link rigid manipulator, which results in a 6-dimensional state-action space and a 11-dimensional policy. In the experiment we discretised the continuous time dynamics into time-steps of $\Delta t = 0.1$ seconds and considered trajectories of 10 seconds in length, which resulted in a planning horizon of $H = 100$. The mean of the initial state distribution was set zero. The elements of the policy parameters $K$, and $m$ were initialised randomly from the interval $[-1, 1]$, while $\pi_\sigma$ was initialised randomly from the interval $[1, 2]$. In the reward function the desired angles of the joints were randomly sampled from the interval $[\pi/4, 3\pi/4]$. All covariance matrices were set to diagonals and the diagonal elements were initialised randomly from the interval $[0, 0.05]$. In all runs of the experiment we gave both algorithms 300 seconds of training time.

The results of the experiment are shown in figure 6 where the normalised total expected reward is plotted against the training time (in seconds). The experiment was repeated 100 times and the plot shows the mean and standard deviation of the results. The plot shows the results for the $Q$-inference algorithm §4 (blue) and the forward-backward inference algorithm of (Hoffman et al., 2009) (red). The dashed line shows that the $Q$-inference algorithm needs only around 35 seconds to obtain the same level of performance as the forward-backward algorithm with 300 seconds of training. Additionally, we can see that after both algorithms have been given the full 300 seconds of training time the forward-backward algorithm only obtains around 70% of the performance compared to that of the $Q$-inference algorithm. Even in this comparatively small experiment the $Q$-inference algorithm therefore significantly outperforms the forward-backward algorithm. The difference in performance of the algorithms can be expected to be even more marked in larger scale

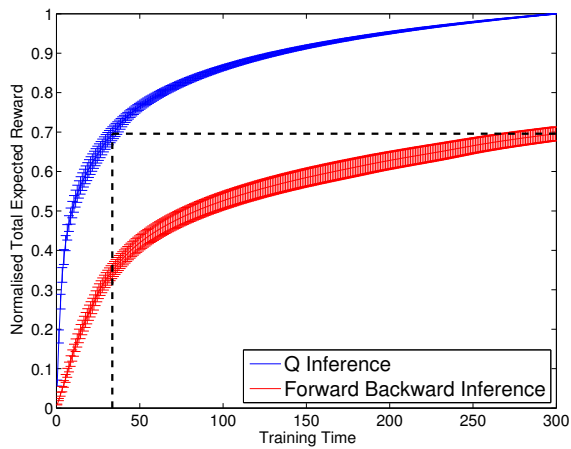

Figure 6: Normalised total expected reward against runtime (in seconds) for the 2-link rigid manipulator problem. The plot shows the results for our continuous $Q$-inference algorithm §4 (blue) and the forward-backward inference algorithm of (Hoffman et al., 2009) (red).

problems and longer planning horizons.

## 6 CONCLUSION

We have presented a new efficient algorithm for performing inference in reward weighted trajectory distributions, which play a prominent role in current state of the art control algorithms like policy gradients and Expectation Maximisation. Our new inference algorithm scales linearly with the planning horizon, whereas the standard forward-backward recursions scales quadratically with the horizon. While we have restricted our attention to Markov decision processes the methods in this paper are readily applicable to other Markovian control problems, such as *partially observable Markov decision processes* (Kaelbling et al., 1998).

Additionally, we have presented a novel algorithm for calculating the sufficient statistics of these distributions in infinite horizon problems, where it is necessary to calculate an infinite number of marginals over a distribution with an infinite number of variables. This has provided an alternative procedure for implementing the EM algorithm in infinite horizon problems.

### References

D. Barber. *Bayesian Reasoning and Machine Learning*. Cambridge University Press, 2011.

P. Dayan and G. E. Hinton. Using Expectation-Maximization for Reinforcement Learning. *Neural Computation*, 9:271–278, 1997.

T. Furmston and D. Barber. Variational Methods for Reinforcement Learning. *AISTATS*, 9(13):241–248, 2010.

M. Hoffman, A. Doucet, N. de Freitas, and A. Jasra. Trans-dimensional MCMC for Bayesian Policy Learning. *NIPS*, 20:665–672, 2008.

M. Hoffman, N. de Freitas, A. Doucet, and J. Peters. An Expectation Maximization Algorithm for Continuous Markov Decision Processes with Arbitrary Rewards. *AISTATS*, 5(12):232–239, 2009.

L. Kaelbling, M. Littman, and A. Cassandra. Planning and Acting in Partially Observable Stochastic Domains. *Artificial Intelligence*, 101:99–134, 1998.

H. Khalil. *Nonlinear Systems*. Prentice Hall, 2001.

J. Kober and J. Peters. Policy search for motor primitives in robotics. *NIPS*, 21:849–856, 2009.

J. Peters and S. Schaal. Policy Gradient Methods for Robotics. *IROS*, 21:2219–2225, 2006.

R. Salakhutdinov, S. Roweis, and Z. Ghahramani. Optimization with EM and Expectation-Conjugate-Gradient. *ICML*, (20):672–679, 2003.

M. Spong, S. Hutchinson, and M. Vidyasagar. *Robot Modelling and Control*. John Wiley & Sons, 2005.

R. Sutton, D. McAllester, S. Singh, and Y. Mansour. Policy Gradient Methods for Reinforcement Learning with Function Approximation. *NIPS*, 13, 2000.

R. S. Sutton and A. G. Barto. *Reinforcement Learning: An Introduction*. MIT Press, 1998.

M. Toussaint. Pros and Cons of truncated Gaussian EP in the context of Approximate Inference Control. *NIPS - Workshop on Probabilistic Approaches for Robotics and Control.*, 21, 2009.

M. Toussaint, S. Harmeling, and A. Storkey. Probabilistic inference for solving (PO)MDPs. Research Report EDI-INF-RR-0934, University of Edinburgh, School of Informatics, 2006.

M. Toussaint, A. Storkey, and S. Harmeling. *Bayesian Time Series Models*, chapter Expectation-Maximization methods for solving (PO)MDPs and optimal control problems. Cambridge University Press, 2011. In press. See userpage.fu-berlin.de/~mtoussai.

M. J. Wainwright and M. I. Jordan. Graphical Models, Exponential Families, and Variational Inference. *Foundations and Trends in Machine Learning*, 1(1-2):1–305, 2008.